\documentclass[twocolumn,showpacs,preprintnumbers,amsmath,amssymb]{revtex4}

\usepackage{graphicx}% Include figure files
\usepackage{dcolumn}% Align table columns on decimal point
\usepackage{bm}% bold math

\begin{document}

\preprint{}

\title{Quantum frequency downconversion experiment}

\author{Hiroki Takesue$^{1,2}$} \email{htakesue@will.brl.ntt.co.jp}
\affiliation{%
$^1$NTT Basic Research Laboratories, NTT Corporation, 3-1 Morinosato Wakamiya, Atsugi, Kanagawa, 243-0198, Japan\\
$^2$CREST, Japan Science and Technology Agency, 4-1-8 Honcho, Kawaguchi, Saitama, 332-0012, Japan
}%

\date{\today}% It is always \today, today,

\begin{abstract}
We report the first quantum frequency downconversion experiment. Using the difference frequency generation process in a periodically poled lithium niobate waveguide, we successfully observed the phase-preserved frequency downconversion of a coherent pulse train with an average photon number per pulse of $<$1, from the 0.7-$\mu$m visible wavelength band to the 1.3-$\mu$m telecom band. We expect this technology to become an important tool for flexible photonic quantum networking, including the realization of quantum repeater systems over optical fiber using atom-photon entanglement sources for the visible wavelength bands. 

\end{abstract}

\maketitle

%\newpage

Quantum frequency conversion \cite{kumar} has been attracting attention as a way to connect photonic quantum information systems with photons of various wavelengths. 
Quantum frequency {\it upconversion} has been employed to convert a telecom-band photon to a visible wavelength photon, so that single photon detector technologies can be utilized for the visible wavelength bands \cite{albota,van,carsten}. These ``upconversion detectors" have been used for several applications including phase-preserved qubit conversion \cite{tanzilli}, quantum key distribution \cite{njp,thew}, and photon-counting optical time domain reflectometry \cite{eleni}.  
Also, an erasure of frequency distinguishability between two single photons was demonstrated using upconversion \cite{takesue}. 

\begin{figure}[thb]

\centerline{\includegraphics[width=\linewidth]{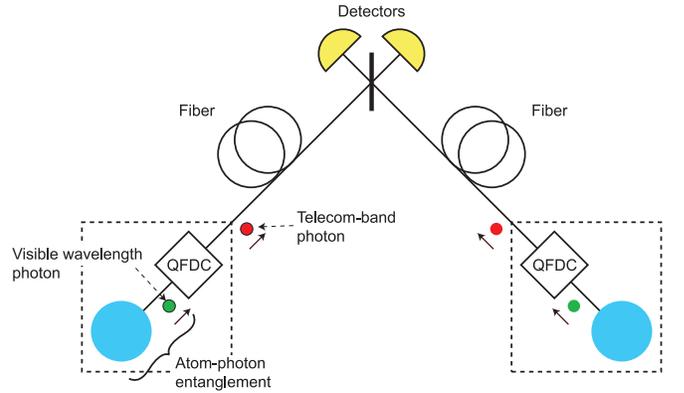}}

\caption{Elementary link for quantum repeater system based on DLCZ or SI schemes. }
\label{system}

\end{figure}

However, to the best of our knowledge, the frequency {\it downconversion} of a single photon has not yet been realized. With quantum frequency downconversion (QFDC), we can add further flexibility to networking quantum information systems over optical fiber. 
For example, we can realize quantum repeater systems \cite{briegel} based on Duan-Lukin-Cirac-Zoller (DLCZ) \cite{dlcz} or Simon-Irvine (SI) schemes \cite{simon} over optical fiber using atom-photon entanglement sources in the visible wavelength bands. 
A configuration for creating entanglement between two distant atomic states, which is called an ``elementary link" in a quantum repeater system based on nested entanglement swapping \cite{briegel}, is shown in Fig. \ref{system}. This link includes two independent atom-photon entanglement sources.  A visible wavelength photon from a source, whose state is entangled with the internal state of the atom, is frequency-converted to the telecom band by using QFDC. The photons from the two sources are transmitted via fibers to an intermediate node where a Bell state measurement (BSM) is performed using linear optics. As a result, we can establish an entanglement between two atomic states when the BSM is successful. 
Since several schemes for generating atom-photon entanglement have already been demonstrated in the visible wavelength band \cite{chou1,monroe,naturepan}, the introduction of QFDC enables us to construct a quantum repeater over optical fiber networks with currently available technologies.

In this paper, we report the first QFDC experiment. Using the difference frequency generation (DFG) process in a periodically poled lithium niobate (PPLN) waveguide, we successfully demonstrated the phase-preserved QFDC of attenuated coherent pulses from the 0.7 $\mu$m band to the 1.3 $\mu$m telecom band.

%%%%%%%%%%%%%%%%%%%%%%%%%%%system

%%%%%%%%%%%%%%%%%%%%%%%%%%%

%theory

Let us briefly describe the theory of the QFDC. We denote the angular frequencies of the pump, the signal and the converted photons as $\omega_p$, $\omega_s$ and $\omega_c$, respectively, which satisfies an energy conservation relationship: $
\omega_p + \omega_c=\omega_s$. 
We input a pump light and a signal photon into a $\chi^{(2)}$ nonlinear medium. When the wavevectors of the signal, pump and converted photons satisfy the phase matching condition, a DFG process involving the three photons occurs in the medium. On the assumption that the pump is strong, the DFG interaction Hamiltonian is given by
\begin{equation}
\hat{H} = i \hbar \chi (a_s a_c^\dagger - H. c.), \label{int}
\end{equation}
where $a_x$ $(x=s,i)$ denotes the annihilation operator, and the subscripts $s$ and $c$ represent the signal and converted photons, respectively. Using this Hamiltonian, we can obtain the operator for the converted photon after a nonlinear interaction time $t$ as
\begin{equation}
a_c (t) = a_c (0) \cos \chi t + a_s (0) \sin \chi t,
\end{equation}
where $a_x (0)$ denotes the annihilation operator for mode $x(=s,c)$ at the input of the nonlinear medium. 
This equation suggests that we can realize phase-preserving state conversion with up to 100\% probability when $\chi t=\pi/2$.

With classical frequency downconversion, we can think of another DFG process, where the energy conservation relationship is expressed as $\omega_p= \omega_c + \omega_s$. When the pump is strong, the interaction Hamiltonian for this process is given by $i \hbar \chi (a_s^\dagger a_c^\dagger - H. c.)$, which corresponds to a parametric amplification process. Therefore, this process inherently generates ``noise photons" through spontaneous parametric downconversion (SPDC), and thus cannot be used for QFDC. 

Even with the process described by the interaction Hamiltonian given by Eq. (\ref{int}), the pump can induce not only QFDC but also SPDC, which can result in the generation of correlated photons whose angular frequencies are $<\omega_p$. We can avoid these photons leaking into the frequency channels for the converted photons by setting $\omega_c > \omega_p$, and this condition is satisfied in the experiment described below. 
%In the experiment explained below, we set the wavelengths of the pump light and converted photons in the 1.5 and the 1.3 $\mu$m bands, respectively, so that we were able to realize low-noise quantum frequency downconversion. 

%setup

%%%%%%%%%%%%%%%%%%%%%%%downconverter

\begin{figure}[thb]

\centerline{\includegraphics[width=\linewidth]{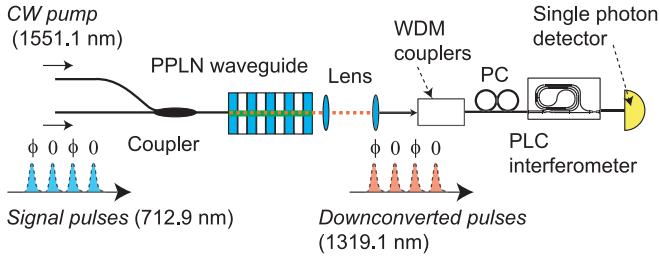}}

\caption{Downconverter setup.}
\label{down}

\end{figure}

%%%%%%%%%%%%%%%%%%%%%%%

Figure \ref{down} shows the setup for the QFDC experiment. As a signal, we prepared a 1 GHz clock sequential coherent pulse train at a wavelength of 712.9 nm, whose phase was modulated alternately with 0 and $\phi$. The scheme for generating this 0.7-$\mu$m pulse train will be described in the next paragraph. The power of the 712.9 nm pulse train was adjusted so that the average photon number per pulse, $\mu$, became less than 1. The pulse train was then input into a fiber-coupled PPLN waveguide module (HC Photonics) for QFDC after being combined with a continuous-wave (CW) pump light with a wavelength of 1551.1 nm emitted from an external cavity diode laser. 
The PPLN waveguide module had a normalized sum-frequency generation (SFG) efficiency of $\sim$200 \%/W. 
The DFG process in the waveguide converted the wavelength of the 712.9-nm signal photons to 1319.1 nm. The photons output from the waveguide were collected by lenses and coupled into a fiber. They were then passed through five 1.3/1.5 $\mu$m wavelength division multiplexing (WDM) couplers to suppress the 1.5-$\mu$m pump light. Then, the converted photons were launched into a 1-bit delayed interferometer fabricated based on planar lightwave circuit (PLC) technology \cite{honjo} after their polarization states were adjusted by a polarization controller (PC). The photons that passed through the interferometer were received by a 4 MHz gated mode single photon detector based on an InGaAs/InP avalanche photodiode. The detection efficiency and dark count probability of the detector were 10\% and 2.6 $\times 10^{-5}$, respectively. 

%%%%%%%%%%%%%%%%%%%%%%%upconverter

\begin{figure}[thb]

\centerline{\includegraphics[width=\linewidth]{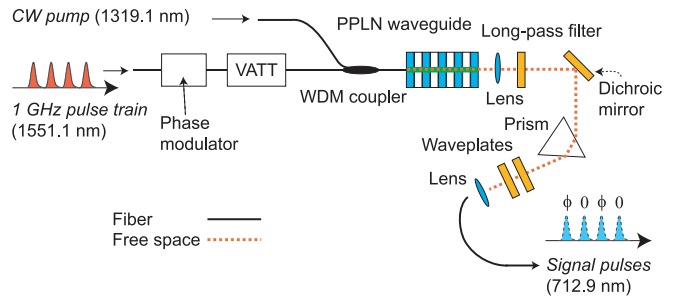}}

\caption{Setup for generating phase-modulated attenuated coherent light at 712.9 nm. }
\label{up}

\end{figure}

%%%%%%%%%%%%%%%%%%%%%%
 
 Figure \ref{up} shows the setup for generating the 712.9-nm, phase-modulated pulses. A 1551.1-nm CW light was modulated into pulses using an optical intensity modulator (not shown). Here, the CW light was obtained from another external cavity laser that was independent of the laser used to generate the 1551.1-nm pump light for the downconverter. The pulse width and interval were 100 ps and 1 ns, respectively. The phases of the 1.5-$\mu$m pulses were modulated alternately with 0 and $\phi$ using an optical phase modulator. The pulses were passed through a variable attenuator (VATT), combined with a 1319.1-nm CW light with a $\sim$1-kHz linewidth from a Nd:YAG laser using a WDM coupler, and launched into a fiber-coupled PPLN waveguide module. 
As a result, a phase-modulated pulse train with the wavelength of 712.9 nm was generated through the SFG process in the PPLN waveguide. 
The light from the waveguide was passed through a long-pass filter to suppress the light generated via the second harmonic generation of the 1.3-$\mu$m pump, and reflected by a dichroic mirror that separated the 0.7-$\mu$m light from the 1.3- and 1.5-$\mu$m lights. 
Then, the 0.7-$\mu$m light was passed through a prism to eliminate the residual 1.3- and 1.5-$\mu$m components, and collimated into a single mode fiber after its polarization had been controlled by waveplates. The $\mu$ value of the 0.7-$\mu$m light was adjusted with a VATT for the 1.5-$\mu$m band installed after the phase modulator.

%result

%%%%%%%%%%%%%%%%%%%%%effandlin
\begin{figure}[bht]

\centerline{\includegraphics[width=\linewidth]{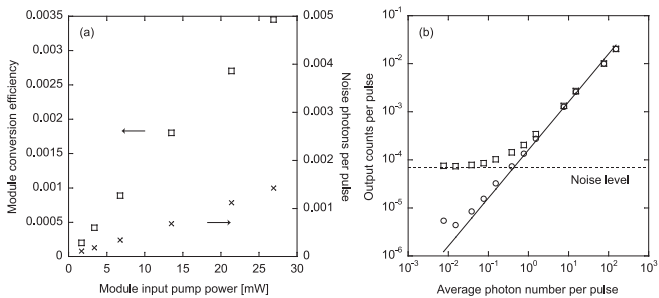}}

\caption{(a) Conversion efficiency as a function of pump power, (b) output count rate per pulse as a function of average input photon number per pulse. }

\label{eff}

\end{figure}

%%%%%%%%%%%%%%%%%%%%%%%%

We first evaluated the downconverter efficiency. 
We eliminated the PLC interferometer and the PC from the setup shown in Fig. \ref{down}, and recorded the count rate observed with the single photon detector. We used a relatively large $\mu$ of 125 to obtain a good signal to noise ratio. The obtained downconversion efficiency as a function of the 1.5-$\mu$m pump power is plotted by squares in Fig. \ref{eff} (a). Here, the efficiency is defined as the number of 1.3-$\mu$m photons at the waveguide output divided by the number of 0.7-$\mu$m photons input into the WDM coupler, and so the WDM coupler loss and the coupling loss between fiber and the waveguide are included. 
We obtained a maximum conversion efficiency of 0.35\% for a pump power of 27 mW. 
%This relatively small efficiency is mainly due to the limited pump power, and thus we can expect a better efficiency simply by improving the 1.5-$\mu$m pump system. 
We also observed a linear increase in the noise count when we increased the pump power with the signal turned off. The estimated number of noise photons per pulse as a function of pump power is shown by the crosses in Fig. \ref{eff} (a). 
The main origin of this noise was the imperfect suppression of the 1.5-$\mu$m pump by the WDM couplers. However, about 20\% of the noise photons probably had other origins, such as a spontaneous Raman scattering process pumped by the 1.5-$\mu$m light in the PPLN waveguide or the fiber pigtail.

To confirm the frequency downconversion of the single photons, we changed the $\mu$ value and measured the output count rate. The count rate per pulse as a function of $\mu$ is shown by the squares in Fig. \ref{eff} (b). Here, we set the pump power at 27 mW, corresponding to a 0.35\% conversion efficiency. 
The solid line shows the linear fitting, indicating that high-fidelity single-photon downconversion was successfully obtained without correction when $\mu$ was $\sim$1 or larger. The dotted line shows the count rate without the signal ($\sim7 \times 10^{-5}$), which includes the counts caused by noise photons and the detector dark counts. Clearly, the observed counts deviated from the linear fitting because of the noise counts when $\mu < 1$. However, when we subtracted the noise counts, the counts fitted very well even when we set $\mu \sim 0.01$, as shown by circles in Fig. \ref{eff} (b). This result suggests that we successfully realized the frequency downconversion of an attenuated pulsed light whose average photon number was much less than 1.

%%%%%%%%%%%%%%%%%%%%%%%%%%%%%%%%%%fringes
\begin{figure}[thb]

\centerline{\includegraphics[width=\linewidth]{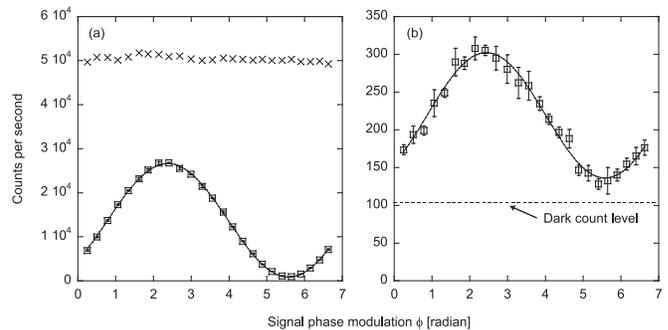}}

\caption{Count rates per second as a function of signal phase modulation $\phi$ (a) $\mu=143$, (b) $\mu=0.7$. }
\label{fringes}

\end{figure}

%%%%%%%%%%%%%%%%%%%%%%%%%%%%%

Next, we confirmed the realization of phase-preserved downconversion at the single-photon level. 
We set the conversion efficiency at 0.35\%, and changed the signal phase modulation value $\phi$. Here, the count rate without the signal was reduced to $\sim3 \times 10^{-5}$, because the PLC interferometer for the 1.3-$\mu$m band suppressed the 1.5-$\mu$m light by $\sim$12 dB, and thus the residual 1.5-$\mu$m noise photons were reduced significantly.  
The detector count rate was observed as a function of $\phi$, while the phase difference induced between the two arms of the interferometer was fixed throughout the measurement. 
The squares in Fig. \ref{fringes} (a) show the observed count rate per second as a function of the signal phase modulation value $\phi$ for $\mu=143$.  
We obtained clear sinusoidal modulation with a 94\% visibility, which demonstrates that the phase modulation applied to the 0.7-$\mu$m signal was transferred to the 1.3-$\mu$m downconverted light. To rule out the possibility that the changes in the count rates were caused by an unwanted intensity modulation, we removed the interferometer and the PC, and undertook the same measurement. 
We found no intensity modulation in the result shown by the crosses in Fig. \ref{fringes} (a). We then undertook another fringe measurement with a reduced $\mu$ value of 0.7. The result is shown in Fig. \ref{fringes} (b), which again shows a clear sinusoidal modulation.   
The fringe visibility was $37.9 \pm 1.1$\%, which was mainly limited by the detector dark count ($2.6 \times 10^{-5}$ per gate, corresponding to 104 Hz with a 4-MHz gate). When we subtracted the detector dark counts, the fringe visibility was $72.1 \pm 2.2$\%, indicating that we can obtain a reasonable fidelity if we use a single photon detector with a smaller dark count rate, such as a superconducting single photon detector (SSPD) \cite{golt}. Thus, we successfully confirmed the realization of phase-preserved QFDC with attenuated pulsed lights where $\mu < 1$.

\begin{figure}[thb]

\centerline{\includegraphics[width=.8\linewidth]{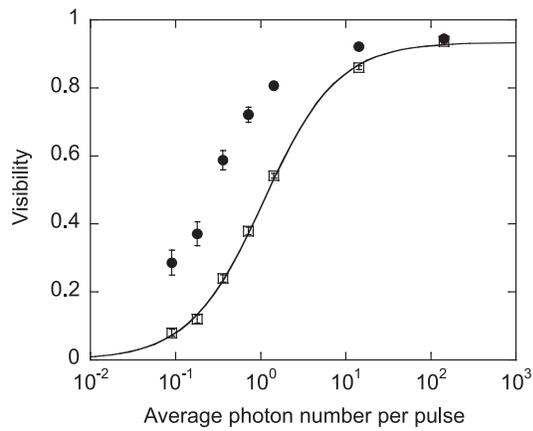}}

\caption{Fringe visibility of the single photon interferences obtained with the downconverted photons as a function of average photon number per pulse. Squares: with raw data, circles: after subtraction of detector dark counts. }
\label{visibility}

\end{figure}

 We then undertook fringe measurements with several $\mu$ values. The obtained fringe visibility as a function of $\mu$ is shown by the squares in Fig. \ref{visibility}. The solid line shows the theoretical curve that takes account of the signal-to-noise ratio degradation caused by the dark counts and noise photons. We managed to observe fringes with $\mu \ge 0.09$. The circles show the visibilities after the subtraction of the detector dark counts. 
 This result suggests that, although the use of a low-noise detector significantly improve the visibility, the residual noise photons limit the fidelity in the present experiment. 
 
 The above result indicates that we need to improve the signal-to-noise ratio of the QDFC process to realize better fidelity. In addition, to apply this technique to quantum information systems, we need to increase the conversion efficiency significantly. The relatively small conversion efficiency observed in the above experiments was mainly attributed to the multimodal characteristics of the PPLN waveguide for the 0.7-$\mu$m light: since the waveguide is not single mode at visible wavelengths, only a fraction of the input 0.7-$\mu$m signal was coupled to the fundamental mode, which effectively reduced the coupling efficiency between the fiber and the waveguide. A solution to this problem is reported in \cite{chou}, where a mode filter and an adiabatic taper are fabricated on a PPLN chip so that a 0.78-$\mu$m light is efficiently coupled to the PPLN waveguide. By utilizing a similar technique, we can expect significant improvements in both conversion efficiency and signal-to-noise ratio. 
 
%%%%%%%%%%%%%%%%%visibility

%%%%%%%%%%%%%%%%%%%%%%%%%%%%%

In conclusion, we successfully demonstrated the QFDC of a single-photon-level light from the 0.7 $\mu$m band to the 1.3 $\mu$m band. We obtained a downconversion efficiency of 0.35\% with a 27 mW pump power, and also confirmed phase-preserved QFDC. 
We expect that this technology will be useful for the flexible networking of quantum information systems over optical fiber. 

The author thanks Q. Zhang, H. Kamada and M. Asobe for fruitful discussions. 

\newpage

%\clearpage

\end{document}